\documentstyle[12pt]{article}
\def\TL{\hfil$\displaystyle{##}$}
\def\TR{$\displaystyle{{}##}$\hfil}
\def\TC{\hfil$\displaystyle{##}$\hfil}
\def\TT{\hbox{##}}
\def\seqalign#1#2{\vcenter{\openup1\jot
  \halign{\strut #1\cr #2 \cr}}}

\def\fixit#1{}

\def\mop#1{\mathop{\rm #1}\nolimits}
\def\tr{\mop{tr}}
\def\Vol{\mop{Vol}}

\def\slashed#1{{\ooalign{\hfil\hfil/\hfil\cr $#1$}}}

\def\href#1#2{#2}  

\def\eqalign#1{\vcenter{\openup1\jot
    \halign{\strut\span\TL & \span\TR\cr #1 \cr
   }}}
\def\lbldef#1#2{\expandafter\gdef\csname #1\endcsname {#2}}
\def\eqn#1#2{\lbldef{#1}{(\ref{#1})}%
\begin{equation} \eqalign{#2} \label{#1} \end{equation}}

\begin{document}
\baselineskip=15.5pt
\pagestyle{plain}
\setcounter{page}{1}

\begin{titlepage}

\begin{flushright}
PUPT-2037 \\
CALT-68-2382 \\
CITUSC/02-014 \\
hep-th/0206098
\end{flushright}
\vfil

\begin{center}
{\huge Non-supersymmetric deformations}
\vskip0.5cm
{\huge of the dual of a confining gauge theory}
\end{center}

\vfil
\begin{center}
{\large Vadim Borokhov$^1$ and Steven S. Gubser$^2$}
\end{center}

$$\seqalign{\span\TL & \span\TT}{
^1 & Lauritsen Laboratory of Physics, 452-48 Caltech, Pasadena, CA  91125  \cr
^2 & Joseph Henry Laboratories, Princeton University, Princeton, NJ 08544
}$$
\vfil

\begin{center}
{\large Abstract}
\end{center}

\noindent 
 We introduce a computational technique for studying
non-supersymmetric deformations of domain wall solutions of interest
in AdS/CFT.  We focus on the Klebanov-Strassler solution, which is
dual to a confining gauge theory.  From an analysis of asymptotics we
find that there are three non-supersymmetric deformations that leave the ten-dimensional
supergravity solution regular and preserve the global bosonic
symmetries of the supersymmetric solution.  Also, we show that there
are no regular near-extremal deformations preserving the global
symmetries, as one might expect from the existence of a gap in the
gauge theory.

\vfil
\begin{flushleft}
June 2002
\end{flushleft}
\end{titlepage}
\newpage
\section{Introduction}
\label{Introduction}

One of the goals of AdS/CFT \cite{juanAdS,gkPol,witHolOne} (see
\cite{MAGOO} for a review) is the study of confinement.  There are
some examples in the literature (for example
\cite{witHolTwo,ks,mnTwo}) where confining gauge theories are dual to
wholly non-singular geometries in supergravity.  We consider it likely
that many other confining theories exist whose duals are singular in
supergravity: examples might include \cite{gDil,gppz}.  In the absence
of a better technical understanding of string theory in Ramond-Ramond
backgrounds, it seems difficult to decide which singularities are
physical (but see proposals in \cite{gNaked,mnOne}).

The purpose of this note is to explore non-singular,
non-supersymmetric deformations of the non-singular supergravity dual
of a confining gauge theory found in \cite{ks}.  Briefly, the gauge
theory is an ${\cal N}=1$ supersymmetric four-dimensional theory with
gauge group $SU(N) \times SU(N+M)$, plus bifundamental matter that can
be summarized by a quiver diagram.  It is thought that the theory goes
through a series of Seiberg dualities that eventually reduce the gauge
group to $SU(M)$, provided $M$ divides $N$.  When the gauge group is
$SU(M)$, the bifundamental matter has disappeared, and the remaining
theory of pure ${\cal N}=1$ supersymmetric glue confines.

The supergravity dual arises from D5-branes wrapped around a shrinking
$S^2$, together with NS-NS three-form field strength and a D3-brane
charge that grows logarithmically with radius.  The geometry is not
asymptotically anti-de Sitter; rather, it can be thought of as nearly
$AdS_5 \times T^{11}$ with a volume for $T^{11}$ that also grows
logarithmically with radius.  Metrically, it is a warped product of
${\bf R}^{3,1}$ with the deformed conifold:
  \eqn{ksAnsatz}{
   ds_{10}^2 = {1 \over \sqrt{h}} (-dt^2 + d\vec{x}^2) + 
    \sqrt{h} ds_6^2 \,,
  }
 where $h$ varies only in the conifold directions, whose Calabi-Yau
metric is $ds_6^2$.  In fact, global symmetries fix a radial direction
in the deformed conifold, and $h$ varies only in this direction.

The solution \ksAnsatz\ is supersymmetric \cite{pg,gSUSY}, and various
properties appropriate to its interpretation as the dual of a
confining gauge theory have been demonstrated in \cite{ks}.  As
mentioned previously, our aim is to consider non-singular,
non-supersymmetric deformations of it.  Such solutions are bound to
have meaning on the gauge theory side, as contrasted to singular
deformations, which might or might not.  A similar analysis has been
carried out \cite{OferPerturb} for the related solution \cite{mnTwo},
where only Neveu-Schwarz three-form flux is present.
Non-supersymmetric generalizations of \cite{mnTwo} were also
considered in \cite{gtv}, where it was found that a discrete series of
disconnected non-supersymmetric vacua exist.  A interpretation on the
gauge theory side was suggested in \cite{ZaffEvans}.

The plan of the paper is as follows.  We introduce our calculation
method in section~\ref{Method}; we summarize the unperturbed solution
in section~\ref{Unperturbed}; we obtain a general form of the renormalized
 mass density of the perturbed solution in
section~\ref{Mass}; and we consider formal and asymptotic properties
of Lorentz-invariant perturbed solutions in sections~\ref{Formal}
and~\ref{Asymptotic}.  At the end of section~\ref{Asymptotic} we reach
the conclusion that there are three regular non-supersymmetric deformations which preserve
the global symmetries of the unperturbed, supersymmetric solution.  In
section~\ref{NonExtremal} we study perturbations that break boost
symmetries in four dimensions but retain all other symmetries, with
the eventual conclusion that there are {\it no} near-extremal
perturbations with regular horizon.  This squares nicely with the
expectation that the field theory has a gap.

\section{The method}
\label{Method}

A generally useful trick for finding supersymmetric solutions is to
parametrize the shape of the compact dimensions by scalars, and then
to show that preserved supersymmetry demands that these scalars obey
the gradient flow equations of some function $W$.  This method goes by
the name of ``attractor equations'' in the study of supersymmetric
black holes in four dimensions \cite{fks}, and it has also been used
extensively in AdS/CFT, with $W$ having the interpretation of a
superpotential---see for example \cite{fgpwOne}, and \cite{Zhukov} for
some remarks on the similarity of the attractor equations and the
superpotential methods for finding domain wall solutions.
``Superpotential methods'' are not in fact restricted to
supersymmetric situations \cite{dfgk,skenderis}: $W$ can be related to
Hamilton's principle function \cite{dBvv}, and only particular forms
(specified by particular choices of integration constants) correspond
to supersymmetric situations.

Our calculational technique for studying perturbations relies on
starting with a solution generated from a known superpotential.
Abstractly, the problem may be cast in the form of supersymmetric
quantum mechanics, with radial variable $u$ playing the role of time.
The radial lagrangian is
  \eqn{Lag}{
   L &= -{1 \over 2} G_{ab} {d\phi^a \over du} 
    {d\phi^b \over du} - V(\phi)  \cr
     &= -{1 \over 2} G_{ab} \left( {d\phi^a \over du} - 
      G^{ac} {\partial W \over \partial\phi^c} \right)
      \left( {d\phi^b \over du} - 
      G^{bd} {\partial W \over \partial\phi^d} \right) +
      {1 \over 2} {dW \over du}
  }
 where
  \eqn{WforV}{
   V = {1 \over 8} G^{ab} {\partial W \over \partial\phi^a}
    {\partial W \over \partial\phi^b} \,.
  }
 (The scalars $\phi^a$ here include components of the metric which
participate in the solution).  The gradient flow equations are
  \eqn{GradientFlow}{
   {d\phi^a \over du} = {1 \over 2} G^{ab} 
    {\partial W \over \partial\phi^b} \,,
  }
 and since the problem is gravitational, there is also a
``zero-energy'' constraint that comes from the $G_{uu}$ Einstein
equation:
  \eqn{Constraint}{
   -{1 \over 2} G_{ab} {d\phi^a \over du} {d\phi^b \over du} + 
     V(\phi) = 0 \,.
  }
 Evidently, a solution to \GradientFlow\ will also solve the equations
of motion for \Lag\ as well as constraint equation \Constraint.  All this is a many-times-told story.  Where we
are introducing something novel is to continue to use the
superpotential as much as possible to study small perturbations to a
solution of \GradientFlow\ that in general satisfy the second order
equations of motion, but not \GradientFlow\ itself.  Let us expand
around a given solution, $\phi_0^a$, to \GradientFlow:
  \eqn{PhiExpand}{
   \phi^a = \phi_0^a + \alpha \bar\phi^a + O(\alpha^2) \,
  }
with a small positive constant $\alpha$.
 It is convenient to introduce further functions 
  \eqn{xiNDef}{
   \xi_a = G_{ab}(\phi_0) \left( {d\bar\phi^b \over du} - 
    N^b{}_d(\phi_0) \bar\phi^d \right) \qquad\hbox{where}\qquad
    N^b{}_a = {1 \over 2} {\partial \over \partial\phi^a} \left(
     G^{bc} {\partial W \over \partial\phi^c} \right) \,.
  }
 Then the linearized equations of motion can be represented as
  \eqn{LinearizedEOMxi}{
   {d\xi_a \over du} + \xi_b N^b{}_a(\phi_0) &= 0\,,
   }
   \eqn{LinearizedEOMphi}{
   {d\bar\phi^a \over du} - N^a{}_b(\phi_0) \bar\phi^b &= 
    G^{ab}(\phi_0) \xi_b \,.
  }
 The constraint can be rephrased as $\xi_a d\phi_0^a/du = 0$.

The set of equations  \LinearizedEOMphi\ is a trivial consequence of the
definition of $\xi_a$.  Equations \LinearizedEOMxi\ may be demonstrated by plugging
the expansion \PhiExpand\ into the equations of motion, in the form
  \eqn{eomForm}{
   &\frac{d}{du}\left( G_{ab}(\phi^{\prime b}-\frac12 G^{bc}\partial_c W)
    \right)+\frac12 \left(\partial_a\partial_b W-(\partial_a G_{bc})G^{cd}
    \partial_d W \right)\left(\phi^{\prime b}-\frac12 G^{bk}\partial_k W 
     \right) \cr&\qquad{}
   -\frac12 (\partial_a G_{bc})\left(\phi^{\prime b}-\frac12 
     G^{bd}\partial_d W \right)\left(\phi^{\prime c}-
      \frac12 G^{ck}\partial_k W \right)=0 \,,
  }
 where primes mean $d/du$.  The constraint can be written as
  \eqn{ConstraintForm}{
   G_{ab}\left(\phi^{\prime a}-\frac12 G^{ac}\partial_c W\right)
    \left(\phi^{\prime b}+\frac12 G^{bd}\partial_d W\right)=0 \,.
  }
 from which $\xi_a d\phi_0^a/du = 0$ easily follows.

Roughly speaking, $\xi_a$ and $\bar\phi^a$ are canonically conjugate. 
It is easy to see that $\xi_a$ describe deformations of the gradient flow equations. Namely, if all $\xi_a$ vanish then
the deformation is supersymmetric. 

Let $\{X\}$ be integration constants parameterizing linearly independent
solutions to \LinearizedEOMxi\ subject to the constraint and a set of 
constants $\{Z\}$ parameterize linearly independent solutions to \LinearizedEOMphi\ with vanishing right-hand side. Since \LinearizedEOMphi\
is a set of nonhomogeneous linear differential equations, general solution for $\bar{\phi}$ is a sum of solutions parametrized by $\{X\}$ and solutions
parameterized by $\{Z\}$. Clearly, $\{X\}$ describe the non-supersymmetric 
deformations, whereas $\{Z\}$ correspond to supersymmetric ones. For example,
deformation generated by $u\to u+\alpha$ is supersymmetric: $\xi_a=0$,
$\bar{\phi}^a=\phi^{\prime a}_0$. As we shall see in section~\ref{Asymptotic},
only specific superpositions of supersymmetric and non-supersymmetric
deformations may prove to be regular.
 
Using \LinearizedEOMxi-\LinearizedEOMphi\ is convenient as a calculational scheme because
one can solve first for the $\xi_a$ through first order equations,
then for $\bar\phi^a$ through more first order equations, rather than
tackling the second order equations directly.  Let us now see the
method in action for a non-trivial example.

\section{Metric Ansatz and Reduced Action }
\label{Unperturbed}

The ansatz of \cite{ks} (see also \cite{pzt}) is the most general one
consistent with the global symmetries of the field theory dual, namely
$SU(2) \times SU(2)$ of flavor and $U(1)_R$.  The metric is
  \eqn{ksMetric}{
   ds^2_{10} &= e^{2p-x+2A}(-dt^2+dx^i dx^i)+
    e^{2p-x+8A}du^2 \cr&\qquad\quad{} + [e^{-6p-x}g^2_5+e^{x+y}(g^2_1+g^2_2)+
    e^{x-y}(g^2_3+g^2_4)] \,,
  }
 and the forms are
  \eqn{ksForms}{
   H_3 &= du\wedge[f^\prime(u)g_1\wedge g_2+k^\prime(u) g_3\wedge g_4)]
   + \frac12[k(u)-f(u)]g_5\wedge(g_1\wedge g_3+g_2\wedge g_4)  \cr
   F_3 &= F(u)g_1\wedge g_2\wedge g_5 +[2P-F(u)]g_3\wedge g_4\wedge g_5+
    F^\prime(u)du\wedge(g_1\wedge g_3+g_2\wedge g_4)  \cr
   F_5 &= {\cal F}_5+{\cal F}^*_5,\quad {\cal F}_5=K(u)g_1\wedge g_2\wedge g_3
    \wedge g_4\wedge g_5  \cr
   &K(u) = k(u)F(u)+f(u)[2P-F(u)] \,,
  }
 where $P$ is a constant.  Explicit expressions for the one-forms
$g_i$ can be found in \cite{ks}.

The eight scalars that will participate in the radial lagrangian are
$\phi^a=(x,y,p,A,f,$ $k,F,\Phi)$.  To obtain the radial lagrangian,
one may start with the type IIB supergravity action (in the
by-now-standard form where $F_5^2$ appears but is set to zero by
imposing self-duality {\it after} obtaining the equations of motion)
and perform the integrals over the angular directions as well as
factoring out the volume of four-dimensional Minkowski spacetime.  The
result is the reduced action: up to an overall factor,
  \eqn{ReducedAction}{
   S[\phi^a]=-\frac{2\Vol_4}{\kappa^2_5}\int\phantom{}du
    \left(-\frac12 G_{ab}{\phi^{\prime a}}{\phi^{\prime b}}-V(\phi)\right),
  }
where 
  \eqn{Kinetic}{
   G_{ab}{\phi^{\prime a}}{\phi^{\prime b}}&=-6A^{\prime 2}+x^{\prime 2} 
    +\frac12 y^{\prime 2}+6p^{\prime 2}  
   \cr &\qquad {}+\frac14\left( \Phi^{\prime 2}+
e^{-\Phi-2x}(e^{-2y}f^{\prime 2}+e^{2y}k^{\prime 2})+2e^{\Phi-2x}F^{\prime 2}
  \right),
  }
and
  \eqn{Potential}{
   V(\phi) &= \frac14 e^{8A-4p-4x}-e^{8A+2p-2x}\cosh{y}+
    \frac14 e^{8p+8A}\sinh^2{y}  \cr&\qquad{}
    +\frac18 e^{8p+8A}\Bigg(\frac12 e^{-\Phi-2x}(f-k)^2+
     e^{\Phi-2x}(e^{-2y}F^2+e^{2y}(2P-F)^2)  \cr&\qquad\qquad\quad{} 
     +e^{-4x}K^2\Bigg) \,.
  }
 The superpotential is (cf.~\cite{pzt})
  \eqn{GiveW}{
   W(\phi)=e^{4A+4p}\cosh{y}+e^{4A-2p-2x}+\frac12 e^{4A+4p-2x}K \,.
  }
 To write down the supersymmetric solution of \cite{ks}, it is
convenient to change radial variables by $du = -e^{-4A-4p} d\tau$.
Then
  \eqn{MetricAgaina}{
   ds^2_{10}=\frac{1}{\sqrt{h(\tau)}}(-dt^2+dx^idx^i)+\sqrt{h(\tau)}ds^2_6 
    \,,
  }
 where $ds^2_6$ is metric of the deformed conifold,
  \eqn{ExplicitDC}{
   ds^2_6&=\frac12 \epsilon^{4/3}N(\tau)\left[ 
    \frac{1}{3N^3(\tau)}(d\tau^2+g^2_5)
    +\cosh^2{\frac{\tau}{2}}(g_3^2+g^2_4)+\sinh^2{\frac{\tau}{2}}(g_1^2+g_2^2)
    \right] \,,  \cr
   &\quad N(\tau)=\frac{(\sinh{2\tau}-2\tau)^{1/3}}{2^{1/3}\sinh{\tau}} \,,
    \quad \epsilon=12^{1/4} \,,  \cr
   h(\tau)&=a\frac{2^{2/3}}{4}\int^\infty_\tau
    \phantom{0}dx\frac{x\coth{x}-1} {\sinh^2{x}}(\sinh{(2x)}-2x)^{1/3} \,, 
    \quad a=\frac{64P^2}{\epsilon^{8/3}} \,,  \cr
   f_0(\tau)&=-\frac{P(\tau\coth{\tau}-1)(\cosh{\tau}-1)}{\sinh{\tau}} \,,\quad
   k_0(\tau)=-\frac{P(\tau\coth{\tau}-1)(\cosh{\tau}+1)}{\sinh{\tau}} \,,  \cr
   F_0(\tau)&=\frac{P(\sinh{\tau}-\tau)}{\sinh{\tau}},\quad \Phi_0=0 \,.
  }

\section{Energy density of the perturbed solution}
\label{Mass}

Let the surface $\partial {\cal M}_u\simeq R^{3,1}\times T^{11}$ 
at fixed $u$ be the boundary
of the interior region ${\cal M}_u$. Consider the boundary metric
$\gamma_{\mu\nu}$, where
  \eqn{BoundaryForm}{
   ds^2=g_{uu}du^2+\gamma_{\mu\nu}\omega^\mu\omega^\nu \,,
    \quad \omega^\mu=(dt,\omega^\alpha) \,,\quad 
    \omega^\alpha=(dx^i,g_1,\dots,g_5) \,.
  }
 Following \cite{BK} we define the quasilocal stress-energy
tensor
  \eqn{QuasiStress}{
   T_{\mu\nu}=\frac{1}{8\pi G}\left(\Theta_{\mu\nu}-\Theta\gamma_{\mu\nu}+
   \frac{2}{\sqrt{|\gamma|}}\frac{\delta S_{ct}}{\delta\gamma^{\mu\nu}} 
    \right) \,,
  }
 where $\Theta_{\mu\nu}$ is the extrinsic curvature
  \eqn{Extrinsic}{
   \Theta_{\mu\nu}=\frac{1}{2\sqrt{g_{uu}}}\partial_u
     \gamma_{\mu\nu} \,,\quad
    \Theta=\Theta_{\mu\nu}\gamma^{\mu\nu} \,,
  }
 and a counterterm action $S_{ct}$ must be chosen to cancel
divergences that appear when $u$ approaches its minimum value ($\tau$ goes to infinity). For a supersymmetric solution the
counterterm is associated with tension of the boundary $\partial{\cal
M}_u$:
  \eqn{Sct}{
   S_{ct}=\int_{\partial{\cal M}_u} 
    dt\wedge dx^i\wedge g_1\wedge\dots\wedge g_5 {\phantom{0}} W(\phi) \,, 
  }
 so that 
  \eqn{deltaSct}{
   \frac{\delta S_{ct}}{\delta\gamma^{\mu\nu}}=
    \frac12 W\gamma_{\mu\nu} \,.
  }
 Let a spacelike surface $\Sigma$ with a metric $\sigma_{\alpha\beta}$ 
be normal to the timelike unit vector
$\frac{1}{N_{\Sigma}}{\partial\over{\partial t}}$: that is, 
  \eqn{NormalSigma}{
   \gamma_{\mu\nu}\omega^\mu\omega^\nu=-N^2_\Sigma dt^2+
   \sigma_{\alpha\beta}\omega^\alpha\omega^\beta \,.
  }
 Then the mass density $M$ is given by
  \eqn{GotMass}{
   M=\int_\Sigma g_1\wedge\dots\wedge g_5 {\phantom{0}}\mu, 
    \quad \mu=\sqrt{|\sigma|}\frac{1}{N_{\Sigma}}T_{tt} \,.
  }
It is possible that this does not vanish even when the perturbed
solution is Lorentz invariant.  In such a case, the mass density
$M$ is interpreted as a contribution to the four-dimensional
cosmological constant.  (But it is consistent that the
four-dimensional geometry is Minkowski space, because the
four-dimensional Newton coupling vanishes for these non-compact
geometries).  If the solution is not Lorentz invariant (as for a
non-extremal deformation), then one should also obtain the spatial
components of $T_{\mu\nu}$. For supersymmetric solution we have
  \eqn{SUSYMass}{
T_{tt}=\frac{3}{8\pi G}e^{-x/2+p-2A}\left( A^\prime-\frac13W\right),  
  \quad \mu=\frac{3}{8\pi G}\left(A^\prime-\frac13W    \right)\,,  
  }
 which are proportional to the gradient flow equation for $A$. Therefore
energy and mass densities of any supersymmetric solution vanish. 

For a non-supersymmetric solution additional counterterms might be needed
\cite{bfs1, bfs2}. If counterterms do not break supersymmetry, 
their contribution to the energy density has the form $O\left[{d\phi^a}/{du}-1/2 G^{ab}{\partial W}/{\partial \phi^b} \right]$. Therefore, for a perturbed solution (\ref{PhiExpand}), we have $T_{tt}$ and $\mu$ of the form $\alpha O[\xi]+O(\alpha^2)$.

\section{Formal Solution}
\label{Formal}

The linearized equations of motion \LinearizedEOMxi-\LinearizedEOMphi\ admit formal
solutions in terms of path ordered exponentials, but this is not very
useful in practice unless the ``connection'' matrix $N^a{}_b(\phi_0)$
is diagonal or triangular---that is, unless the equations decouple, or
can be solved iteratively.  This occurs only partially for our case,
as we shall develop in this section by examining the explicit form of
the equations and extracting the simplest combinations of them that we
can.

First, let us use the radial variable $\tau$m ($du=-e^{-4A_0-4p_0}d\tau$), so that
\LinearizedEOMxi-\LinearizedEOMphi\ become
  \eqn{tauEOM}{
   \dot{\xi}_a+\xi_b M^b{}_a(\phi_0)&=0  \cr
   \dot{\bar{\phi}^a}-M^a{}_b(\phi_0)\bar{\phi}^b&=
    -e^{-4A_0-4p_0}G^{ab}(\phi_0) \xi_b \,,  \cr
   M^b{}_a &= -e^{-4A_0-4p_0}N^b{}_a \,.
  }
 Given a vector, like $\xi_a$, we will have frequent occasion to use
the shorthand notation $\xi_{3+4} = \xi_3 + \xi_4$, or $\xi_{5-6} =
\xi_5 - \xi_6$.  

First let us deal with the first line of \tauEOM.  The vector
  \eqn{NullV}{
   V = (V^a) = 
   \left( 3,0,-1,1,\frac{3}{P}[2fP+F(k-f)],\frac{3}{P}[2fP+F(k-f)],0,0
    \right)
  }
 is annihilated by $M^a{}_b$: that is, $M^a{}_b V^b = 0$.  It follows
that
  \eqn{etadot}{
   \dot{\eta}+\frac{3}{P}[2f_0 P+F_0(k_0-f_0)]\dot{\xi}_{5+6}=0 \,,
  }
 where $\eta=3\xi_1-\xi_3+\xi_4$ and $\xi_{5+6} = \xi_5+\xi_6$ as per
our shorthand notation explained above.  We also find
  \eqn{xidot}{
   \dot{\xi}_{5+6}+\frac13 e^{-2x_0}P\eta=0 \,,
  }
 and now we can solve
  \eqn{etaxiSolve}{
   \eta(\tau) &= X_\eta \exp{\left(\int^\tau_{\tau_0} d\tau\phantom{0}e^{-2x_0}
    [2f_0P+F_0(k_0-f_0)]\right)}  \cr
   \xi_{5+6}(\tau) &= X_{5+6}-\frac{P}{3}\int^\tau_{\tau_0} d\tau
    \phantom{0}e^{-2x_0}\eta(\tau) \,,
  }
 where $X_\eta$ and $X_{5+6}$ are integration constants.  It can be
further shown that
  \eqn{MoreXi}{
   \dot{\xi}_{5-6}+\xi_7&=\frac13 e^{-2x_0}(F_0-P)\eta \,,  \cr
   \dot{\xi_7}+\cosh{(2y_0)}\xi_{5-6}&=
    \frac16 e^{-2x_0}(f_0-k_0)\eta-\sinh{(2y_0)}\xi_{5+6} \,.
  }
 Once $\xi_{5-6}$ and $\xi_7$ are determined, we can find $\xi_8$ using
  \eqn{XiEight}{
   \dot{\xi}_8=e^{2y_0}(2P-F_0)\xi_5+e^{-2y_0}F_0\xi_6+\frac12(f_0-k_0)\xi_7 
    \,.
  }
 Using the constraint $\xi_a \dot\phi_0^a = 0$, together with the
solution for $\eta$, we may solve for $\xi_3$ and $\xi_4$, and then we
find two coupled linear differential equations for $\xi_1$ and
$\xi_2$. This completes a formal solution of
$\dot\xi_a + \xi_b M^b{}_a = 0$ together with $\xi_a \dot\phi_0^a =
0$.

We have still to solve the second equation in \tauEOM, that is,
$\dot{\bar\phi^a} - M^a{}_b(\phi_0) \bar\phi^b = -e^{-4A_0-4p_0}
G^{ab}(\phi_0) \xi_b$.  For $\bar{\Phi}$ we have
  \eqn{BarPhiDot}{
   \dot{\bar{\Phi}}=-4e^{-4A_0-4p_0}\xi_8 \,.
  }
Functions $Y(\tau)=\bar{x}(\tau)-3\bar{A}(\tau)$, $\bar{\phi}_{3+4}(\tau)=\bar{p}(\tau)
+\bar{A}(\tau)$ and $\bar{y}$
satisfy 
  \eqn{ThreePhis}{
   \dot{\bar{\phi}}_{3+4}+
      e^{-2x_0-6p_0}\bar{\phi}_{3+4}+e^{-2x_0-6p_0}Y &=
    \frac16 e^{-4p_0-4A_0}(\xi_4-\xi_3)  \cr
   \dot{Y}+4\cosh{(y_0)}\bar{\phi}_{3+4}+\sinh{(y_0)}\bar{y} &=
     -e^{-4p_0-4A_0}(\xi_1+\frac12\xi_4)  \cr
   \dot{\bar{y}}+\cosh{(y_0)}\bar{y}+4\sinh{(y_0)}\bar{\phi}_{3+4} &=
     -2e^{-4p_0-4A_0}\xi_2 \,.
  }
 For $\bar{\phi}_{5-6}=\bar{f}-\bar{k}$ and $\bar{F}$ one obtains
  \eqn{TwoMorePhis}{
   \dot{\bar{F}}-\frac12\bar{\phi}_{5-6} &=
     \frac12(f_0-k_0)(4\bar{\phi}_{3+4}-
     \bar{\Phi})-2e^{2x_0-4p_0-4A_0}\xi_7,  \cr
   \dot{\bar{\phi}}_{5-6}-2\cosh{(2y_0)}\bar{F} &= 
     4(\sinh{(2y_0)}F_0 -e^{2y_0}P)\bar{y}  \cr&\hskip-1in
     +2(\cosh{(2y_0)}F_0-e^{2y_0}P)(4\bar{\phi}_{3+4}+\bar{\Phi})
     +8e^{2x_0-4p_0-4A_0} \sinh{(2y_0)}(\xi_5-\xi_6) \,.
  }
 A function $\bar{\phi}_{5+6}=\bar{f}+\bar{k}$ satisfies
  \eqn{OneMorePhi}{
   \dot{\bar{\phi}}_{5+6} &= -4(e^{2y_0}P+\cosh{(2y_0)}F_0)\bar{y}+
    2(\sinh{(2y_0)}F_0-e^{2y_0}P)(4\bar{\phi}_{3+4}+\bar{\Phi})  \cr&\qquad{}
    +2\sinh{(2y_0)}\bar{F}-8e^{2x_0-4p_0-4A_0}\cosh{(2y_0)}(\xi_5+\xi_6) \,.
  }
Finally, for $\bar{A}$ we have
  \eqn{LastPhi}{
   &\dot{\bar{A}}+e^{-2x_0}(2f_0P+F_0(k_0-f_0))\bar{A} =
    \frac13 \sinh{(y_0)}\bar{y}+
     \frac23[2e^{-2x_0}f_0P+e^{-2x_0}F_0(k_0-f_0)  \cr&\qquad{}
    -e^{-2x_0-6p_0}+2\cosh{(y_0)}] \bar{\phi}_{3+4}-
    \frac13 e^{-2x_0}[2f_0P+F_0(k_0-f_0)+2e^{-6p_0}]Y  \cr&\qquad{}
    +\frac16 e^{-2x_0}(2P-F_0)\bar{f}
    +\frac16 e^{-2x_0}F_0\bar{k}
    +\frac16 e^{-2x_0} (k_0-f_0)\bar{F} +\frac16 e^{-4p_0-4A_0}\xi_4 \,.
  }
 This completes a formal solution of the linearized equations~\tauEOM.

\section{Asymptotic Solutions}
\label{Asymptotic}

Unfortunately, even the partially decoupled equations that we found in
the previous section do not appear to admit analytic solutions.  What
we can do, however, is to give a complete treatment of the asymptotics
in the regions $\tau \to 0$ (in the interior, where we may require
that the solution be completely regular) and at $\tau \to \infty$,
where we can sensibly require that the asymptotics is unchanged at the
leading order from the supersymmetric solution and that the energy
density is finite.

We will use the algorithm described in the previous section to find
asymptotic form of the perturbations $\bar{\phi}$ in the regions
$\tau\to 0$ and $\tau\to\infty$.

\medskip\noindent
{\bf Large $\tau$:} Observing that
 \eqn{LeadingTermsInf}{\seqalign{
  \span\TL & \span\TR & \quad\span\TT & \span\TL & \span\TR}{
   (\frac{\partial}{\partial\phi} W)(\phi_0) &\sim \frac{e^{4/3\tau}}{\tau}
    & for $\phi=(x,p,f,k)$\,, & 
   (\frac{\partial}{\partial A} W)(\phi_0) &\sim e^{4/3\tau} \,, \cr
   (\frac{\partial}{\partial\phi} W)(\phi_0) &\sim e^{\tau/3} 
    & for $\phi=(y,F)$\,, &
   G^{aa}(\phi_0)e^{-4p_0-4A_0} &\sim \tau e^{-4/3\tau}\,,
  }}
 we conclude that we must solve for $\xi$ up to the terms $\sim 1/\tau$
at infinity; find $\bar{x}$, $\bar{f}$, and $ \bar{k}$ 
up to $\tau e^{-4/3\tau}$; $\bar{p}$ and $\bar{A}$ up to $e^{-4/3\tau}$;
and $\bar{y}$ and $\bar{F}$ up to $e^{-\tau/3}$.  The solution for
$\xi(\tau)$ has the following form:
\eqn{XiInf}{\seqalign{\span\TL & \span\TR &
 \hskip-0.5in\span\TR}{
\xi_1  &=X^\infty_1 e^{2\tau}-4X^\infty_1\tau-4X^\infty_1+6PX^\infty_{5-6}-P
X^\infty_{5+6}+2X^\infty_2+O(e^{-\tau}) \,,\span\cr
\xi_2 &=2(PX^\infty_{5-6}-X^\infty_1)e^\tau\tau
+X^\infty_2 e^\tau
+O(e^{-\tau}) \,, & \qquad\qquad
\xi_3 =3X^\infty_1e^{2\tau}-12X^\infty_1\tau+O(e^{-\tau}) \,,\cr
\xi_4 &=-18PX^\infty_{5-6}+3PX^\infty_{5+6}-6X^\infty_2+12X^\infty_1+ O(e^{-\tau}) \,, \span\cr
\xi_5  &=\frac12 X^\infty_{5-6}e^\tau+\frac12 X^\infty_{5+6}+O(e^{-\tau}) \,,
&
\xi_6 =-\frac12 X^\infty_{5-6}e^\tau+\frac12 X^\infty_{5+6}+O(e^{-\tau})\,,
\cr
\xi_7 &=-X^\infty_{5-6}e^\tau+O(e^{-\tau}) \,, &
\xi_8 =(X^\infty_{5+6}-2X^\infty_{5-6})P\tau+X^\infty_8+O(e^{-\tau}) \,.
}}
where $X^\infty_1, X^\infty_2, X^\infty_{5-6}, X^\infty_{5+6}, X^\infty_8$ are
integration constants. The other integration constants parametrize 
solutions which are $ O(e^{-\tau})$.
For the dilaton deformation we have
\eqn{DInf}{
\bar{\Phi} &=Z^\infty_8+3^{1/3} 4PX^\infty_{5+6}e^{-4/3\tau}(3+4\tau)-
3^{1/3} 8 PX^\infty_{5-6}e^{-4/3\tau}(3+4\tau)\cr&\qquad{}
 +3^{1/3} 16 X^\infty_8 e^{-4/3\tau}+\dots \,,
}
 where $Z^\infty_8$ is an integration constant. To keep $\Phi(\infty)$
fixed, we set \eqn{DilConstant}{Z^\infty_8=0\,.}  For $\bar{f}$ and
$\bar{k}$ we have
 \eqn{fANDkInf}{
\bar{f} &=\frac12 Z^\infty_{5+6}+\frac12 Z^\infty_7 e^{-\tau}-\frac12 P
Z^\infty_2 e^{-\tau}(1+2\tau)+3^{1/3}12 P X^\infty_2(e^{-\tau/3}-4\tau 
e^{-4/3\tau}) \cr&\qquad{}
+3^{1/3}P^2 X^\infty_{5-6} (-96 \tau^2 e^{-4/3\tau}-96\tau 
e^{-4/3\tau}-90e^{-\tau/3})+ 3^{1/3} 24P^2 X^\infty_{5+6}\tau e^{-4/3\tau}
\cr&\qquad{}
+P Z^\infty_Y(e^{4/3\tau}+O(\tau e^{\tau/3}))
+Z^\infty_{5-6}(\frac12 
e^\tau-2\tau)+3^{1/3} 6 PX^\infty_1 e^{2/3\tau}+..., \cr
\bar{k} &=\frac12 Z^\infty_{5+6}-\frac12 Z^\infty_7 e^{-\tau}+\frac12 P
Z^\infty_2 e^{-\tau}(1+2\tau)- 3^{1/3}12 P X^\infty_2(e^{-\tau/3}+4\tau 
e^{-4/3\tau}) \cr & \qquad{}
+3^{1/3}P^2 X^\infty_{5-6} (-96 \tau^2 e^{-4/3\tau}-96\tau 
e^{-4/3\tau}+90e^{-\tau/3})+ 3^{1/3}24 P^2 X^\infty_{5+6}\tau e^{-4/3\tau}
\cr & \qquad{}
+P Z^\infty_Y(e^{4/3\tau}+O(\tau e^{\tau/3})) 
+Z^\infty_{5-6}(-\frac12 
e^\tau-2\tau)+3^{1/3} 6PX^\infty_1 e^{2/3\tau}+\dots \,,
}
where $Z^\infty_{5-6}$, $Z^\infty_{5+6}$, $Z^\infty_Y$, $Z^\infty_7$, 
and $Z^\infty_2$ are integration constants. Although $k_0(\tau)$ and $f_0(\tau)$  are divergent at infinity, the divergence is linear in $\tau$. Therefore, we require vanishing of the exponentially divergent terms in \fANDkInf: 
\eqn{RegInfin}{
X^\infty_1=Z^\infty_Y=Z^\infty_{5-6}=0 \,.
}
For the rest of deformations $\bar{\phi}$ we have
\eqn{MorePhiInf}{\seqalign{\span\TL & \span\TR &\qquad 
    \span\TL & \span\TR}{
\bar{x} &=-3^{1/3} 48 P X^\infty_{5-6}\tau e^{-4/3\tau}+
Z^\infty_4(3\tau^{-4/3} e^{4/3\tau}+O(e^{-2/3\tau}))\span\span\cr
&-\frac{3^{1/3}}{4^{4/3}P}
Z^\infty_{5+6}\tau^{-4/3}e^{4/3\tau}\Gamma(\frac43,\frac43\tau)+\dots\,,
\span\span\cr
\bar{y} &=3^{1/3} 16 PX^\infty_{5-6} e^{-\tau/3}(3-2\tau)-
3^{1/3}16X^\infty_2 e^{-\tau/3}+\dots \,, \span\span\cr
\bar{p} &=-3^{1/3}PX^\infty_{5-6} e^{-4/3\tau}(10+16\tau)+Z^\infty_4(-\tau^{-4/3}
e^{4/3\tau}+O(e^{-2/3\tau})) \span\span\cr
& +3^{1/3}4e^{-4/3\tau}(-2X^\infty_2+PX^\infty_{5+6})
+\frac{1}{3^{2/3} 4^{4/3}P} Z^\infty_{5+6}\tau^{-4/3}e^{4/3\tau}\Gamma
(\frac43,\frac43\tau)+\dots\,, \cr
\bar{A} &= 3^{1/3}PX^\infty_{5-6} e^{-4/3\tau}(67-8\tau)+Z^\infty_4
(\tau^{-4/3}e^{4/3\tau}+O(e^{-2/3\tau}))\span\span\cr
& -\frac{1}{3^{2/3} 4^{4/3}P} Z^\infty_{5+6}\tau^{-4/3}e^{4/3\tau}\Gamma
(\frac43,\frac43\tau)-3^{1/3}6X^\infty_2 
e^{-4/3\tau}-3^{4/3}PX^\infty_{5+6}e^{-4/3\tau}+\dots\,,\cr
\bar{F} &=3^{1/3}P^2 X^\infty_{5-6}e^{-\tau/3}(126-48\tau)-
3^{1/3}36PX^\infty_2e^{-\tau/3}+\dots\,, 
}}
where $Z^\infty_4$ is a constant and $\Gamma(\frac43,z)$ is the 
incomplete gamma function\footnote{
$\Gamma(a,z)=\int_z^\infty\phantom{0}t^{a-1}e^{-t} dt$.}
\eqn{GammaExpan}{
\Gamma(\frac43,z)=e^{-z}\left(z^{1/3}+\frac13 z^{-2/3}+O\left(\frac{1}{z^{5/3}}\right)
\right),\quad
z\to\infty\,.
}
Requiring deformed solution to have the same asymptotic as \ExplicitDC\ we  
have \eqn{RegularityInfinity}{Z^\infty_4=0 \,.}

\medskip\noindent
{\bf Small $\tau$:} We observe that
\eqn{LeadTermZero}{
G^{aa}(\phi_0)e^{-4p_0-4A_0} &\sim 1/\tau^2\,,\quad 
\hbox{for $a=(1,2,3,4,6,8)$\,,} \cr
G^{77}(\phi_0)e^{-4p_0-4A_0} &= O(1)\,, \quad
G^{55}(\phi_0)e^{-4p_0-4A_0} \sim \tau^2\,, \cr
h(\tau) &= h_0+h_2\tau^2+O(\tau^4)\,,
}
where $h_0$ and $h_2$ are constants. 
For $\xi(\tau)$ the leading asymptotics are 
\eqn{XiZero}{\seqalign{\span\TL & \span\TR &\qquad 
    \span\TL & \span\TR}{
\xi_1 &=\frac13 X^0_\eta-\frac{32}{27}\frac{P^2X^0_\eta}{h_0}+\frac23\frac{X^0_\eta h_2}{h_0}-\frac25P X^0_7+2X^0_1+O(\tau^2)\,,\span\span\cr
\xi_2  &=\frac23\frac{X^0_\eta h_2}{h_0}-\frac{32}{27}\frac{P^2X^0_\eta}{h_0}
-\frac{16}{15}P X^0_7+2X^0_1+O(\tau^2) \,, &
\xi_3 &=-\frac23 X^0_\eta+O(\tau^2)\,, \cr
\xi_4 &=-\frac23 X^0_\eta+\frac{32}{9}\frac{P^2X^0_\eta}{h_0}
-2\frac{X^0_\eta h_2}{h_0}
+\frac65 P X^0_7
-6X^0_1+O(\tau^2)\,, \span\span\cr
\xi_5 &=\frac{X^0_7}{2\tau}+\frac{4PX^0_\eta}{3h_0\tau}+X^0_{5+6}+O(\tau) \,, 
&
\xi_6 &=-\frac{X^0_7}{2\tau}-\frac{PX^0_\eta\tau}{9h_0}\, \cr
&+\frac{1}{12}X^0_7\tau
+O(\tau^2)\,, &
\xi_7 &=\frac{X^0_7}{\tau^2}+\frac16X^0_7+O(\tau)\,, \cr
\xi_8 &=X^0_8+\frac{8P^2X^0_\eta\tau^2}{27h_0}+\frac{2}{15}PX^0_7\tau^2
+O(\tau^3)\,,
}}
where $X^0_\eta$, $X^0_1$, $X^0_{5+6}$, $X^0_7$ and $X^0_8$ are constants. 
The dilaton deformation is given by
\eqn{DilZero}{
\bar{\Phi}=\frac{16}{\tau}X^0_8+Z^0_8+O(\tau)\,,
}
where $Z^0_8$ is a constant. Regularity of $\bar{\Phi}$ at small $\tau$
implies $X^0_8=0$. We have 
$${\seqalign{\span\TL & \span\TR &\qquad 
    \span\TL & \span\TR}{
\bar{x} &=\frac{2}{\tau^3}Z^0_2+\frac{4}{3\tau}X^0_\eta
\left(1-\frac{20P^2}{9h_0}+\frac{2h_2}{h_0}-\frac{8P^2h_2}{3h_0^2}
+\frac{128P^4}{27h_0^2}\right)+\frac{8}{\tau}X^0_1\left(1-\frac{4P^2}{3h_0}
\right)-\frac{4P}{h_0\tau}Z^0_{5-6} \,\span\span\cr
&+\frac{8P}{\tau}X^0_7\left(-\frac15+\frac{2h_2}{h_0}
+\frac{32P^2}{45h_0
}\right)+\frac{88P^2}{45h_0\tau}Z^0_2-\frac{2P}{h_0\tau}
Z^0_{5+6}-Z^0_{3+4}+3Z^0_4+O(\tau\log{\tau})\,, \span\span\cr
\bar{y} &=-\frac{2}{\tau^3}Z^0_2+\frac{16}{3\tau}X^0_\eta\left(\frac{h_2}{h_0}
-\frac{16P^2}{3h_0} \right)+\frac{16}{\tau}X^0_1-\frac{128P}{15\tau}X^0_7
+\frac{7}{3\tau}Z^0_2+4Z^0_{3+4}+O(\tau\log{\tau})\,, \span\span\cr   
\bar{p} &=\frac{1}{\tau^3}Z^0_2-\frac{4}{9\tau}X^0_\eta\left(1+\frac{4P^2}{3h_0} -\frac{8P^2h_2}{3h_0^2}+\frac{128P^4}{27h_0^2} \right)+\frac{32P^2}{9h_0\tau}X^0_1 -\frac{2}{15\tau}Z^0_2\left( 1+\frac{44P^2}{9h_0}\right)\, \span\span\cr
& -\frac{4P^2}{3\tau}X^0_7\left(1+\frac{4h_2}{h_0}+\frac{64P^2}{45h_0}
\right)
 +\frac{4P}{3h_0\tau}Z^0_{5-6}+\frac{2P}{3h_0\tau}Z^0_{5+6}+Z^0_{3+4}-Z^0_4+O(\tau\log{\tau})\,, \span\span\cr
\bar{A} &=\frac{2}{15\tau}Z^0_2\left(1+\frac{44P^2}{9h_0} \right)\
+\frac{4}{9\tau}X^0_\eta\left(1-\frac{4P^2}{h_0}+\frac{3h_2}{h_0}
-\frac{8P^2h_2}{3h_0^2}+\frac{128P^4}{27h_0^2}\right)
-\frac{4P}{3h_0\tau}Z^0_{5-6}\, \span\span\cr
&+\frac{4}{\tau}X^0_1\left( 1-\frac{8P^2}{9h_0}\right) +\frac{P}{\tau}X^0_7\left( -\frac45+\frac{16h_2}{3h_0}+\frac{256P^2}{135h_0}
\right) -\frac{2P}{3h_0\tau}Z^0_{5+6}+Z^0_4+O(\tau)\,,
\span\span
  }}$$\eqn{PhiZero}{\seqalign{\span\TL & \span\TR &\qquad \span\TL & \span\TR}{
\bar{f} &=\frac12 Z^0_{5-6}+\frac12 Z^0_{5+6}+O(\tau^2)\,, &
\bar{k} &=\frac{56P}{5\tau^2}Z^0_2+\frac{16P}{3\tau^2}X^0_\eta
\left(1-\frac{2h_2}{h_0}+\frac{32P^2}{9h_0} \right)\, \cr
& -\frac{32P^2}{\tau^2}X^0_1+\frac{16}{\tau^2}X^0_7\left(3h_2+\frac{16P^2}{15}
\right)-\frac{6}{\tau^2}Z^0_{5-6}-\frac12 Z^0_{5-6}
+\frac12 Z^0_{5+6}+O(\tau\log{\tau})\,, \span\span\cr
 \bar{F} &=\frac{64P}{15\tau}Z^0_2
+\frac{8P}{3\tau}X^0_\eta\left( 1-\frac{2h_2}{h_0}+\frac{32P^2}{9h_0} \right)
-\frac{16P}{\tau}X^0_1+\frac{2}{\tau}X^0_7\left(h_0+12h_2
+\frac{64P^2}{15}\right)\, \span\span\cr
&-\frac{3}{\tau}Z^0_{5-6}+O(\tau)\,,\span\span
}}
 where $Z^0_2$, $Z^0_{3+4}$, $Z^0_4$, $Z^0_{5-6}$, and $Z^0_{5+6}$ 
are constants.  The perturbation of the metric is small if
\eqn{RegMetric}{
2\bar{p}-\bar{x}+2\bar{A}\le0,\quad -6\bar{p}-\bar{x}\le0,\quad \bar{x}+\bar{y}\le0,\quad \bar{x}-\bar{y}\le0.
}
 Regularity of $\bar{f}$, $\bar{k}$, $\bar{F}$ together with
\RegMetric\ implies
 \eqn{RegForms}{\seqalign{\span\TL & \span\TR &\qquad 
    \span\TL & \span\TR}{
Z^0_2 &=X^0_7=0\,, & Z^0_{5-6} &=\frac{8P}{9}X^0_\eta\,,\cr
X^0_1 &=\left(\frac{16P^2}{27h_0}-\frac{h_2}{3h_0}\right)X^0_\eta \,, & 
Z^0_{5+6} &=\left(\frac{2h_0}{3P}-\frac{8P}{9}\right)X^0_\eta \,.
}}

\medskip\noindent 
 {\bf The space of regular deformations:} As far as we can see, it
would require numerics on the coupled system~\tauEOM\ to describe any
particular linearized deformation of the basic solution~\ExplicitDC.
However, having found the asymptotics both at large and small $\tau$,
we can meaningfully inquire how many independent deformations there
are, consistent with the basic ansatz, \ksAnsatz\ and \ksForms.  There
are fifteen integration constants for the system~\tauEOM: seven $\{X\}$
and eight $\{Z\}$.  At large
$\tau$ we have constrained five of them: one of $\{X\}$ and four of $\{Z\}$, through dilaton asymptotics
\DilConstant, regularity conditions \RegInfin\ and
\RegularityInfinity. At small $\tau$ regularity of \DilZero\ and \RegForms\
have constrained three of $\{X\}$ and three of $\{Z\}$. Two of these
$\{Z\}$ are related to $\{X\}$ through \RegForms.  We have not shown that these are 
independent constraints, but on grounds of genericity we expect that 
this is so. 
According to our discussion in section~\ref{Method} we conclude that  there are three unconstrained real parameters, 
corresponding to three regular non-supersymmetric deformations. 

This is not an easy result to understand directly from the field
theory, partly because of the effects of strong coupling.  The field
content, in ${\cal N}=1$ superfield language, is
 \eqn{AllFields}{\seqalign{
   \span\TC\qquad & \span\TC\quad & \span\TC\quad & 
     \span\TC\quad & \span\TC}{
    & SU(N) & SU(N+M) & SU(2)_A & SU(2)_B  \cr
  V_1 & \hbox{adj} & 1 & 1 & 1  \cr
  V_2 & 1 & \hbox{adj} & 1 & 1  \cr
  A & \overline{N} & N+M & 2 & 1  \cr
  B & N & \overline{N+M} & 1 & 2
 }}
 where we have also indicated the quantum numbers under the gauge
symmetry $SU(N) \times SU(N+M)$ and the global symmetry $SU(2)_A
\times SU(2)_B$.  The first two lines show real superfields, while the
second two show chiral superfields.  Deformations of the supergravity
solution of the type we consider should correspond to adding gauge
theory operators to the lagrangian which are invariant under all the
symmetries in \AllFields.  One difficulty with this point of view is
that the theory is strongly coupled, in two ways: first, there is the
usual large 't~Hooft coupling limit associated with the geometry being
smooth on the string scale; and second, the theory is close to a
Leigh-Strassler fixed point with a quartic superpotential, so it's not
clear that a simple lagrangian suffices to describe the physics.  If
we set this aside and write down a list of the gauge operators which
are singlets under the symmetries in \AllFields\ and are relevant or
marginal based on the dimensions one obtains from the Leigh-Strassler
fixed point, there are considerably more than three: the list includes
  \eqn{ManyOps}{\seqalign{\span\TC}{
   \tr [F_{\mu\nu}^{(1)}]^2 \,, \quad
   \tr [F_{\mu\nu}^{(2)}]^2 \,, \quad
   \tr \lambda_{(1)}^2 \,, \quad
   \tr \lambda_{(2)}^2 \,, \quad
   \tr \lambda_{(1)} \slashed{D} \lambda_{(1)} \,, \quad
   \tr \lambda_{(2)} \slashed{D} \lambda_{(2)} \,, \cr
   \tr |a_i|^2 \,, \quad \tr |b_i|^2 \,, \quad
   \tr |D_\mu a_i|^2 \,, \quad \tr |D_\mu b_i|^2  \cr
   \tr \bar\psi_{a_i} \psi_{a_i} \,, \quad
   \tr \bar\psi_{b_i} \psi_{b_i} \,, \quad
   \tr \bar\psi_{a_i} \slashed{D} \psi_{a_i} \,, \quad
   \tr \bar\psi_{b_i} \slashed{D} \psi_{b_i} \,, \cr
   \epsilon^{ik} \epsilon^{jl} \tr a_i b_j a_k b_l \,, \quad
   \epsilon^{ik} \epsilon^{jl} \tr \psi_{a_i} \psi_{b_j} a_k b_l \,,
  }}
 where we have employed some obvious notation: $\lambda_{(1)}$ is the
gaugino component of $V_1$, $a_i$ is the scalar component of $A_i$,
$\psi_{a_i}$ is the fermion component of $A_i$, and the Roman indices
$i,j,\ldots$ are labels for the fundamentals of $SU(2)_A$ or
$SU(2)_B$.  In writing down the list \ManyOps, we have incorporated
also an R-symmetry constraint: $U(1)_R$ is broken to a ${\bf Z}_2$
which should be respected by operators dual to the supergravity
deformations that we have considered.  Note that the fields $a_i$ and
$b_j$ both have R-charge $1/2$ \cite{ks}.

Two mechanisms cut down the list of operators that should appear in
\ManyOps.  First (and trivially), we constrained $\Phi(\infty)$, so
only the difference $\tr [F_{\mu\nu}^{(1)}]^2 - \tr
[F_{\mu\nu}^{(2)}]^2$ should be allowed, not the sum.  Second, some
operators unrelated to chiral primaries must be expected to acquire
large anomalous dimensions and to be dual to excited string states.
It is not entirely clear even at the Leigh-Strassler fixed point which
operators these are, but this is something that might be elucidated in
the future.  For a field theory with a moduli space of vacua, one
might expect that a third mechanism would cut down the list of allowed
deformations of the lagrangian, namely the requirement of a stable
vacuum.  However, as in the case of \cite{OferPerturb}, the fact that
the vacuum of the unperturbed theory is isolated tells us that any
small deformation of the lagrangian should still have a stable vacuum.
The upshot is that once one sorts out which operators correspond to
supergravity modes as opposed to excited string states, it should be
possible to identify the three allowed deformations on the gauge
theory side.  We hope to return to this problem in the future.

\section{Non-Extremal Deformation}
\label{NonExtremal}

Because the gauge theory confines, we might reasonably expect that
there are no near-extremal generalizations of the
solution~\ExplicitDC\ with regular horizons.  To understand the
reasoning, let us first note that solutions with regular horizons do
exist with the same asymptotics at infinity \cite{ghkt}, but the
horizon entropy of these solutions scales as $N^2$, and their gauge
theory interpretation is in terms of a high temperature, deconfined
phase with restored chiral symmetry.  It shouldn't be possible for
such a phase to persist arbitrarily close to zero energy density and
temperature; hence the conclusion that {\it near-extremal}
generalizations with regular horizons should not exist.

To explore this explicitly in our setup of linearized supergravity
perturbations, let us make the following ansatz for the metric:
 \eqn{NExtrMetric}{
ds^2_{10}=e^{2p-x+2A+2z}(-e^{-8z}dt^2+dx^i dx^i)+
e^{2p-x+8A}du^2\, \cr
+[e^{-6p-x}g^2_5+e^{x+y}(g^2_1+g^2_2)+
e^{x-y}(g^2_3+g^2_4)]\,.
}
Let  $\phi^A=(\phi^a,z)$ with $\phi^a=(x,y,p,A,f,k,F,\Phi)$.
The reduced action has the form
\eqn{RedActionNExtr}{
S[\phi^A]=-\frac{2\Vol_4}{k^2_5}\int\phantom{}du(-\frac12 G_{AB}{\phi^{\prime A}}{\phi^{\prime B}}-V(\phi))\,,
}
where 
\eqn{KineticNExtr}{
G_{AB}{\phi^{\prime A}}{\phi^{\prime B}}=6z^{\prime 2}+G_{ab}\phi^{\prime^a}
\phi^{\prime b}\,,
}
and the potential $V(\phi)$, the superpotential $W(\phi)$, and the
metric $G_{ab}$ are the same as in Section \ref{Unperturbed}:
\eqn{PotentEn}{
V=\frac18 G^{AB}\partial_A W\partial_B W=\frac18 G^{ab}\partial_a W
\partial_b W\,. 
}
The constraint equation has the following form:
\eqn{ConstraintNExtr}{
G_{ab}\left(\phi^{\prime a}-\frac12 G^{ac}\partial_c W\right)
\left(\phi^{\prime b}+\frac12 G^{bd}\partial_d W\right)+6=0\,.
}
 The  supersymmetric solution satisfies
\eqn{SuSySol}{
\phi^{\prime A}=\frac12 G^{AB}\partial_B W\,.
}
 We notice that both the target-space metric $G_{AB}$ and
superpotential $W$ are independent of $z$, which implies that $z$ is
constant in the supersymmetric solution and can be absorbed into
redefinition of $t$ and $x^i$.

Let 
\eqn{DeformPhi}{
\phi^A=\phi^A_0+\alpha\hat{\phi}^A+\alpha^2\tilde{\phi}^{A}+O(\alpha^3)\,,
}
where $\alpha$ is a deformation parameter and $z_0=0$. Equation 
$\delta S/\delta z=0$ implies $z^{\prime\prime}=0$. To study the non-extremal deformations we
choose
\eqn{Z}{
z=\alpha \bar{z},\quad \bar{z}=u\,.
}
Substitution of \DeformPhi\ into the equations of motion and constraint 
gives (to first order in $\alpha$)
\eqn{DifEqForg}{
\frac{d}{du}g_a+g_b N^b_a(\phi_0)=0\,, \quad  g_a \phi^{\prime a}_0=0\,,
}
where
\eqn{Defineg}{
g_a=G_{ab}(\phi_0) (\hat{\phi}^{\prime b}-N^b_d(\phi_0) 
\hat{\phi^d}),\quad N^b_a=\frac12\partial_a
(G^{bc}\partial_c W)\,.
}
Equations \DifEqForg\ and  \Defineg\ do not involve $z$ and 
correspond to the extremal deformation of the supersymmetric
 solution. To study non-extremal deformation we set
$\hat{\phi^a}=0$. 
For $\tilde{\phi^a}$ we have
\eqn{DifEqZeta}{
\frac{d}{du}\zeta_a+\zeta_b N^b_a(\phi_0)=0,\quad
  \zeta_a \phi^{\prime a}_0+3=0\,,
}
where
\eqn{DefineZeta}{
\zeta_a=G_{ab}(\phi_0) (\tilde{\phi}^{\prime b}-N^b_d(\phi_0) \tilde{\phi^d})\,.
}
 Solutions for $\zeta$ and $\tilde{\phi}$ can be written in terms of
the general solution for $\xi$ and $\bar{\phi}$. For small $\tau$ we have
  \eqn{zetaForms}{\seqalign{\span\TL & \span\TR &\qquad 
    \span\TL & \span\TR}{
   \zeta_1 &= -\frac{12}{\tau}-4\tau+\xi_1+O(\tau^2) \,, &
   \zeta_2 &= -\frac{6}{\tau}-\frac{43}{5}\tau+\xi_2+O(\tau^2) \,,  \cr
   \zeta_3 &= \frac{12}{\tau}-\frac{28}{5}\tau+\xi_3+O(\tau^2) \,, &
   \zeta_4 &= \frac{48}{\tau}+\frac{32}{5}\tau+\xi_4+O(\tau^2) \,,  \cr
   \zeta_5 &= \xi_5+O(\tau) \,, &
   \zeta_6 &= \xi_6+O(\tau^2) \,,  \cr
   \zeta_7 &= \xi_7+O(\tau) \,, &
   \zeta_8 &=\xi_8+O(\tau^3) \,,  \cr
   \tilde{x} &= -\frac{16}{\tau^2}+64\log{\tau}-
    \frac{424}{15}+\bar{x}+O(\tau\log{\tau}) \,,\span\span\cr
   \tilde{y} &= -\frac{16}{\tau^2}+128\log{\tau}-
    \frac{824}{15}+\bar{y}+O(\tau\log{\tau}) \,,\span\span\cr
   \tilde{p} &= \frac{8}{3\tau^2}-\frac{128P^2}{3h_0}\log{\tau}+
    \frac{256P^2}{27h_0}+\bar{p}+O(\tau\log{\tau}) \,,\span\span\cr
   \tilde{A} &= -\frac{32}{3\tau^2}+\left(32+
    \frac{128P^2}{3h_0}\right)\log{\tau}-\frac{256P^2}{27h_0}+
    \bar{A}+O(\tau) \,,\span\span\cr
   \tilde{f} &= 32P\tau+\bar{f}+O(\tau^2) \,, &
   \tilde{k} &= -\frac{64P}{3\tau}+\bar{k}+O(\tau\log{\tau}) \,,  \cr
   \tilde{F} &= -\frac{16}{3}P+\bar{F}+O(\tau) \,, &
   \tilde{\Phi} &= \bar{\Phi}+O(\tau) \,.
  }}
 We observe that for any values of the integration constants, a
function $\tilde{k}$ is infinite at $\tau=0$. Therefore, we conclude
that non-extremal linearized deformations \DeformPhi\ are singular at
the apex of deformed conifold.  Thus we reach the desired conclusion
that near-extremal solutions with regular horizons do not exist.

\section*{Acknowledgments}

V.B.\ would like to thank Arkadas Ozakin, Xinkai Wu, and Takuya Okuda
for useful discussions.  The work of V.B.\ was supported in part by
the DOE under grant DE-FG03-92ER40701.  The work of S.S.G.\ was
supported in part by the DOE under grants DE-FG03-92ER40701 and
DE-FG02-91ER40671 and through an Outstanding Junior Investigator
Award.  S.S.G.\ gratefully acknowledges Caltech's support while this
work was carried out.

\bibliography{gap}
\bibliographystyle{ssg}

\end{document}